\documentclass{elsart}
\usepackage{graphicx}% Include figure files
\usepackage{dcolumn}% Align table columns on decimal point
\usepackage{amssymb}
\journal{Physica A}
\begin{document}
\begin{frontmatter}

\title{Application of Unit Time Block Entropy to Fetal Distress Heart Rate}

\author{U.C. Lee \corauthref{cor}}
\corauth[cor]{Corresponding author.}
\ead{uclee@postech.ac.kr}
\author{S.Kim}
\ead{swan@postech.ac.kr}
\address{Asia Pacific Center for Theoretical Physics \& Nonlinear Complex Systems Laboratory, National Core Research Center on System Biodynamics, Department of Physics, Pohang University of Science and Technology, Pohang, 790-784, Korea}

\date{\today}

\begin{abstract}
Recently, multiple time scale characteristics of heart dynamics have received much attention for distinguishing healthy and pathologic cardiac systems. Despite structural peculiarities of the fetal cardiovascular system, the fetal heart rate(FHR) displays multiple time scale characteristics similar to the adult heart rate due to the autorhythmicity of its different oscillatory tissues and its interaction with other neural controllers. In this paper, we investigate the event and time scale characteristics of the normal and two pathologic fetal heart rate groups with the help of the new measure, called the Unit Time Block Entropy(UTBE), which approximates the entropy at each event and time scale based on symbolic dynamics. This method enables us to match the measurement time and the number of words between fetal heart rate data sets simultaneously. We find that in the small event scale and the large time scale, the normal fetus and the two pathologic fetus are completely distinguished. We also find that in the large event scale and the small time scale, the presumed distress fetus and the acidotic distress fetus are significantly distinguished. 
\end{abstract}

\end{frontmatter}

%%%%%%%%%%%%%%%%%%%%%%%%%%%%%%%%%%%%%%%%%%%%%%%%%%%%%%%%%%%%%%%%%%%%%

\section{\label{sec:level1}Introduction}

Fetal distress is generally used to describe the lack of the oxygen of the fetus, which may result in damage or death if not reversed or the fetus is delivered immediately. Thus the emergent action by discriminating the fetal distress is an important issue in the obstetrics. In particular, it is important to discriminate the normal Fetal Heart Rate(FHR) and two types of fetal distress groups(the presumed distress and the acidotic distress) and reduce the wrong diagnosis rate in which about 65\% of the presumed distress fetus(not serious but in need of careful monitoring) are diagnosed as the acidotic distress fetus(serious, needing an emergent action), experiencing a useless surgical operation. In this paper, we try to discriminate the normal and pathologic fetal heart rate group with a robust and reliable method.\\
The cardiovascular system of the fetus is a complex system. The complex signal, the heart rate, from the complex system contains enormous information about the various functions. The incoming information, on which various heart activities are projected, can not be linearly separated into each function. But we may be able to estimate the signal from a pathologic fetal heart which contains information on the weakness or a complete loss of a particular function. Based on this concept, multiple time scale characteristic of heart dynamics have received much attention. As a quantitative method, the Multi-Scale Entropy, which quantifies multi-time scale complexity in the heart rate, was introduced and widely applied to the classification between normal and pathologic groups and also different age groups.
It was found that in a specific time scale region, the normal and the pathologic adult heart rate groups and the different age effects on the heart rhythms are significantly distinguished~\cite{Altimiras1999,Havlin1999,Costa2002,Chialvo2002,Costa2002_2,Costa2003,Costa2005}.\\ 
The multiple-scale characteristic in the fetal heart rate is due to the autorhythmicity of its different oscillatory tissues, its interaction with other neural controller and the maternal circulatory system and other neural or hormonal activities, which have a wide range of time scales from secondly to yearly\cite{Wood1999}.\\
In our previous work on the multiple-scale analysis of heart dynamics, we extended the analysis with the time scale to both the event and time scales\cite{Lee2005}. We found that the event scale in heart dynamics plays more important role than the time scale in classifying the normal and pathologic adult groups. In this paper, therefore, we will investigate characteristic event or time scales of heart dynamics of the normal and the fetal distress group in order to determine the criteria for classifying the normal and the fetal distress groups.\\
Previous works on the fetal heart rate were based on the various nonlinear measures such as the multi-scale entropy, approximate entropy, power spectral density and detrended fluctuation analysis~\cite{Magenes2002,Magenes2003,Signorini2003}.
They were able to significantly differentiate the mean of normal and pathologic FHR groups but their classification performance was poor, which prevent practical applications of these methods. 
In this paper, we investigate the typical scale structure of each FHR group by scanning both the event and time scale regions in order to find an appropriate scale region for classifying the normal and the pathologic fetal groups in an optimal way.\\
We introduce a new analysis method, called the Unit Time Block Entropy(UTBE), which scans both the event and time scale regions of the heart rate based on symbolic dynamics to find the characteristic scale region of the normal and the pathologic heart rate groups\cite{Lee2005}. This method matches the unevenly sampled RR interval data length and the measurement time of the heart rate data set simultaneously, where the RR interval means the time duration between consecutive R waves of the electrocardiogram(ECG). In most previous studies the number of RR interval sequences of all data set are fixed in spite of a large variability in the measurement time\cite{Costa2002,Chialvo2002,Costa2002_2,Costa2003,Costa2005,Magenes2002}. Using the UTBE method, we can directly compare the entropy of data sets without any ambiguity caused by the nonstationarity and noise effect that inevitably appears in the data set with different data length or measurement time. We find that the normal and two pathologic FHR groups are reliably discriminated.\\
In Section 2, we will introduce the normal and pathologic FHR data set and their linear properties. In Section 3, the new analysis method, called Unit Time Block Entropy(UTBE), is introduced and applied to the fetal heart rate data set. In Section 4, we show that the normal, the presumed distress and the acidotic distress FHR groups can be discriminated through the UTBE method in some characteristic scale regions. Finally, we end with the conclusion.
%%%%%%%%%%%%%%%%%%%%%%%%%%%%%%%%%%%%%%%%%%%%%%%%%%%%%%%%%%%%%%
\section{The normal and pathologic fetal heart rates and their linear properties}
%%%%%%%%%%%%%%%%%%%%%%%%%%%%%%%%%%%%%%%%%%%%%%%%%%%%%%%%%%%%%%%%%%%%%%%
%FIGURE1
\begin{figure}
\begin{center}
\includegraphics[keepaspectratio,width=0.6\columnwidth]{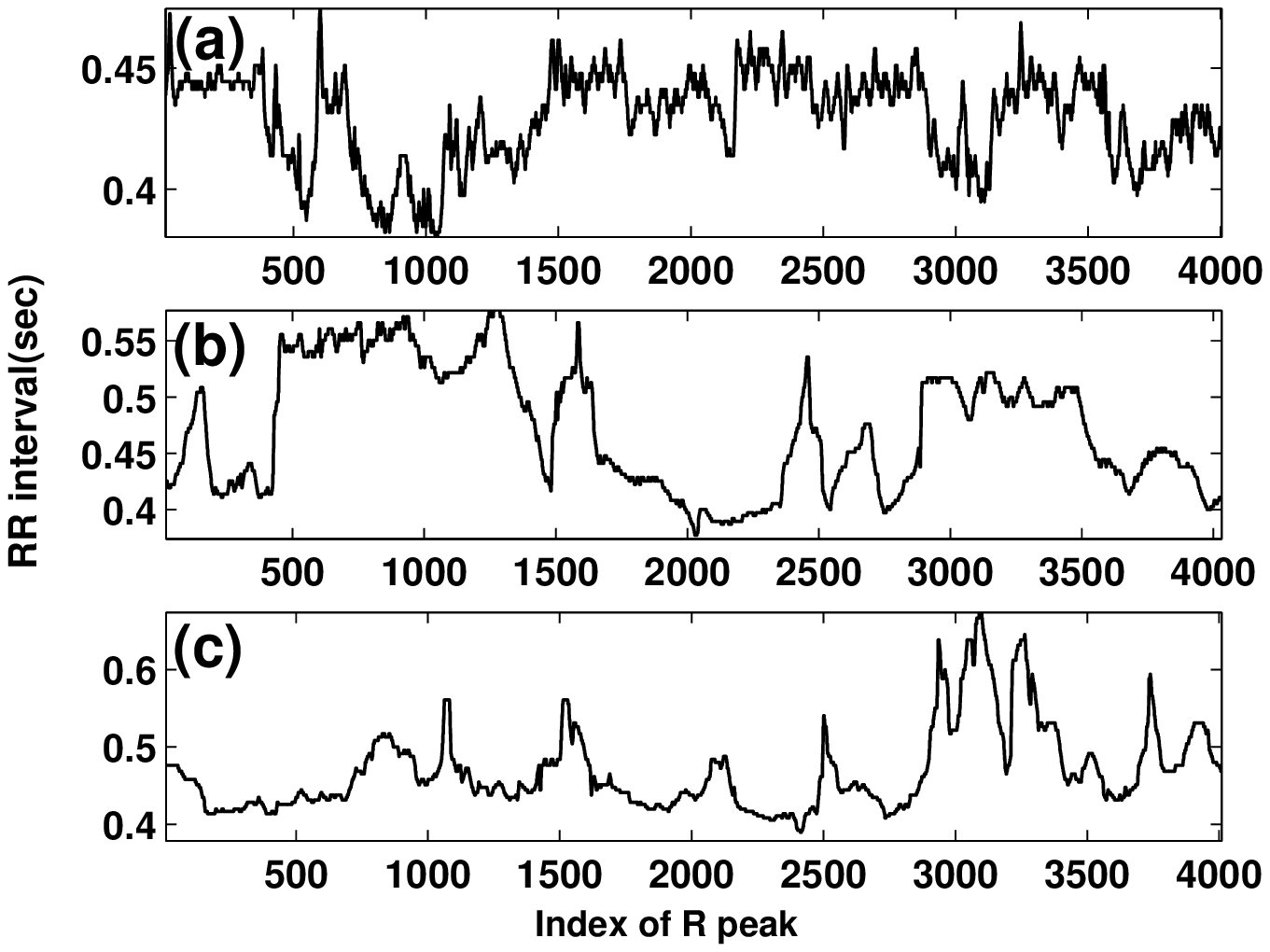}\\
\includegraphics[keepaspectratio,width=0.4\columnwidth]{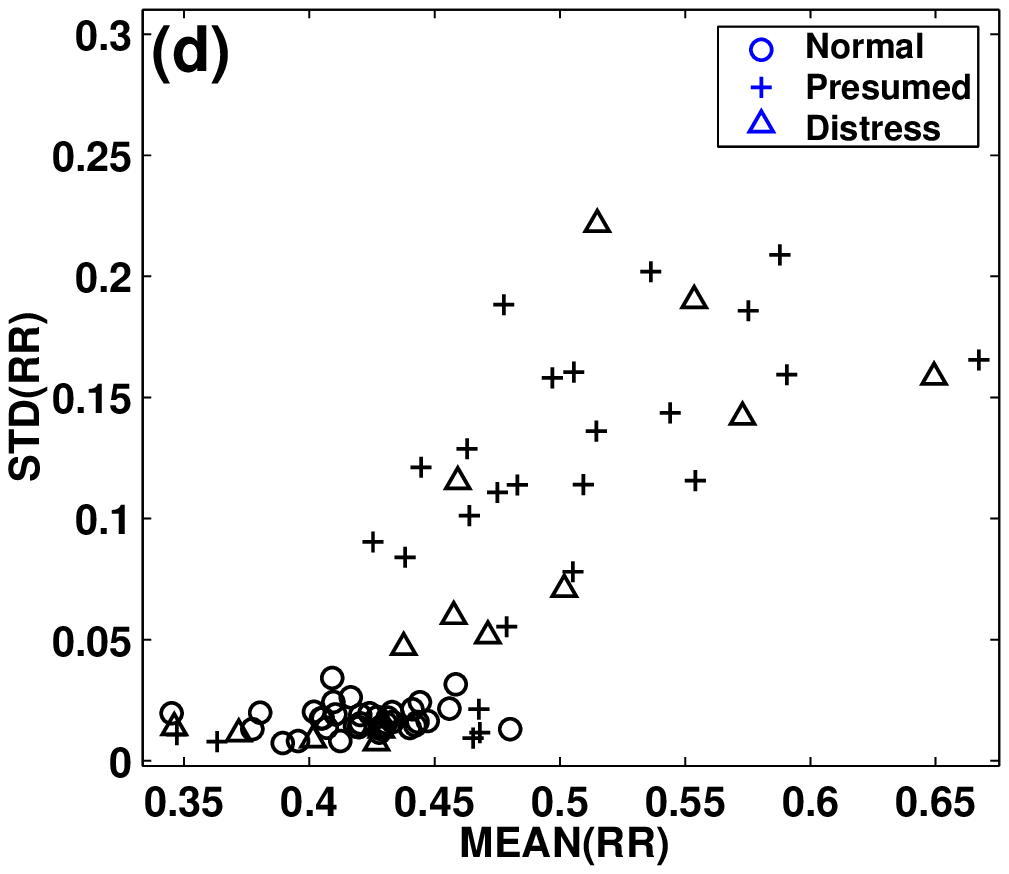}
\includegraphics[keepaspectratio,width=0.4\columnwidth]{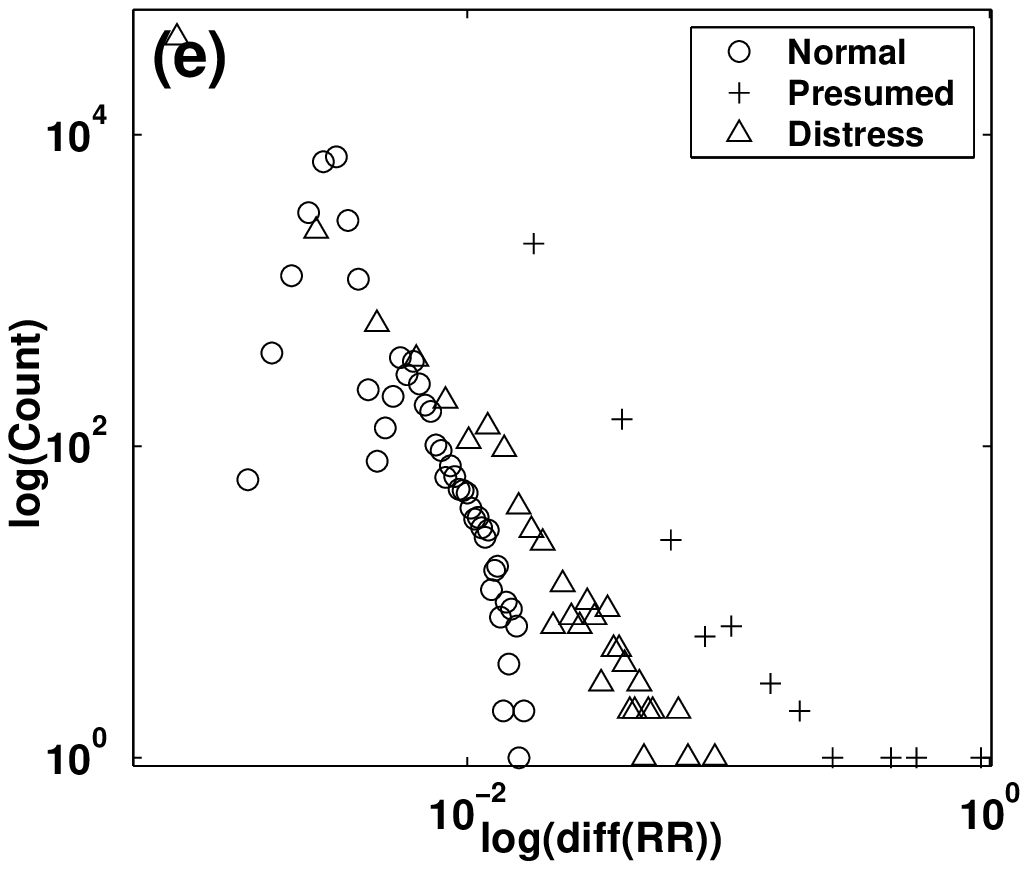}\\
\end{center}
\caption{\label{fig:fig3} The RR interval time of three fetuses. (a) A normal fetus. (b) A presumed fetus. (c) A distress fetus. (d)  The mean and standard deviation of normal, presumed and distress RR interval sequences in units of seconds. (e) The log-log plot of RR interval acceleration. The log-log distribution of RR acceleration shows a power-law distribution.}
\end{figure}
%%%%%%%%%%%%%%%%%%%%%%%%%%%%%%%%%%%%%%%%%%%%%%%%%%%%%%%%%%%%%%%%%%%%%%%
The fetal heart rate is acquired from 77 pregnant women who were placed under computerized electronic fetal monitoring during the ante partum and intra partum periods. The fetal heart rate is digitized with the data received with the Corometrics 150 model(Corometrics, Connecticut, USA) through the Catholic computer-assisted obstetric diagnosis system(CCAOD; DoBe Tech, Seoul, Korea). The computerized electronic fetal heart rate monitoring was done on the fetal heart rate data two hours before the delivery, from each pregnant woman without any missing data. First, 77 pregnant women are divided into two groups; 36 women into the normal fetus group, who showed the normal heart rate tracing, and 41 women with the abnormal fetal heart rate tracing(severe variable, late deceleration, bradycardia, or tachycardia, and delivery by caesarean section). The heart rate tracing has been done using as a standard criteria to determine normal or pathologic states of a fetus. After an immediate delibery by a caesarean section, the umbilical artery $pHs$ of the fetus is examined. Then, 41 women were further divided into the presumed distress group with 26 women whose umbilical artery $pHs$ was higher than 7.15, and the acidotic distress group with 15 women whose umbilical artery $pHs$ was lower than 7.15 and the base excess was lower than $-8mM/l$. By this umbilical artery $pHs$ test we can retrospectively know whether a fetus was in dangerous situation or not. In this work, the wrong diagnosis rate is $26/41(63.4\%)$, 26 women undertaken a useless surgical operation. The improvement of this wrong diagnosis rate through the presurgical classification of different groups based on the FHR data is the main goal of this work. \\
In order to treat errors in the measuring equipment and ectopic beats we selected the interval that did not have any missing data, and manually treated the ectopic beats using a linear interpolation, which is less than $0.1\%$ of all heart beats selected. We replaced the ectopic beats with resampled data around the vicinity of $2Hz$ using the average sampling frequency of each heart rate sequence. The resampling, which is usually used for even sampling from the unevenly measured heart beat, is restrictively applied for replacing ectopic beats. However, this method inevitably produces a number of unexpected artificial effects on the original RR sequences. For example, it can cause the distortion of the short term correlation of heart rate data\cite{Gomes2002}. Thus a new method is developed and applied to overcome these problems as in the following section. \\ 
%%%%%%%%%%%%%%%%%%%%%%%%%%%%%%%%%%%%%%%%%%%%%%%%%%%%%%%%%%
\subsection{Linear Properties}
Before applying the nonlinear method, we investigate conventional linear properties such as the mean and the standard deviation and its student t-test between healthy and pathologic groups(normal, presumed distress and acidotic distress).
In Fig.1(a)-(c), the RR interval time of three fetuses are presented. The normal fetus shows a typical characteristic of the small, fast variations, while the pathologic fetus shows relatively slower, larger variations. In Fig.1(d) and (e), linear properties such as the mean, the standard deviation and the log-log distribution of RR acceleration from the normal, the presumed distress and the acidotic distress groups are presented. The result of the t-test on the mean and the standard deviation distributions shows that three FHR groups are distinguishable by these two linear properties, resulting in significant p-values for the mean and the standard deviation[p-value(for the mean/the standard deviation): the normal and the presumed distress($10^{-5}/8 \times 10^{-11}$), the normal and the acidotic distress($0.018/3\times 10^{-5}$), the presumed distress and the acidotic distress(0.28/0.903)]. Although their significant difference in the mean and the standard deviation, their classification performance based on the sensitivity and specificity is poor. In the next section, we will compare the linear properties with the nonlinear ones obtained from the UTBE and the multi-scale entropy method. 
In Fig.1(e), the RR interval acceleration shows a power-law distribution. This suggests that the RR interval acceleration rather than RR interval has a nonlinear characteristics which can be best utilized in classifying the different FHR groups.
%%%%%%%%%%%%%%%%%%%%%%%%%%%%%%%%%%%%%%%%%%%%%%%%%%%%%%%%%%%%%%%%%%
\section{Unit Time Block Entropy(UTBE)}

\begin{figure}
\begin{center}
\includegraphics[keepaspectratio,width=0.55\columnwidth]{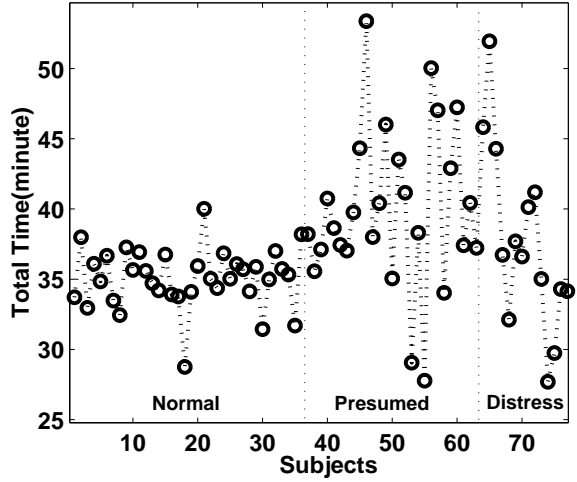}
\end{center}
\caption{\label{fig:fig1} The variation of measurement time after fixing the number of RR interval sequence in healthy and pathologic FHR groups. The mean and standard deviation of $77$ subjects is $37.25\pm 5.04$(min), it shows a large variation.}
\end{figure}

The investigation of linear properties such as the mean and the standard deviation of the RR sequence shows that the heart dynamics of the normal fetus group is faster than that of the pathologic fetus group, leading to a significant difference in the mean. However, the total measurement time shows a large variability for the 77 subjects studied. In Fig.\ref{fig:fig1}, the mean and the standard deviation of the measurement time of all RR interval sequence is $37.25\pm5.04(min)$ and the time difference between the shortest one and the longest one is $25.69(min)$. In treating the RR interval sequence, we typically face this contradictory situation. If we try to match the number of RR interval sequence in advance, the measurement time for all RR interval sequences becomes different. If we try to match the measurement time in advance, the number of RR interval sequence becomes different. This naturally occurs for unevenly sampled data such as the RR interval sequence. In order to solve this problem, we used the Unit Time Block Entropy(UTBE), which simultaneously matches both the measurement time and the number of RR interval sequence for all subjects.
%%%%%%%%%%%%%%%%%%%%%%%%%%%%%%%%%%%%%%%
%FIGURE1
The UTBE estimates the entropy of the symbolic sequence composed at the specific event and time scale of a heart beat sequence. Since it matches both the measurement time and the number of words of the alphabet from the RR sequence, we can make a direct comparison of all involved FHR groups reliably.\\
First, we define a word sequence from the RR interval sequence. To construct a word, we use a unit time windowing method, in which each word is constructed in a given unit time window, so the number of symbols involved in each word can be different between windows.\\
The RR interval sequence is given by $TI=\{x_1,x_2,\dots,x_i,\dots,x_n\}$ and the RR interval acceleration is by $\Delta TI=\{t_1,t_2,\dots,t_i,\dots,t_{n-1}\},~~t_i=x_{i+1}- x_{i}$. 
A word is defined as follows
\begin{eqnarray}
    w_i &=& \{s_{i}s_{i+1}\dots  s_{i+j}\dots  s_{i+n(i)} \},\\
    s_{i+j} &=& \left\{ \begin{array}{ll}
    0~,& ~~ \textrm{if $|t_{i+j}| \leq \tau $} \\
    1~,& ~~ \textrm{if $|t_{i+j}| > \tau $}\end{array} \right.
\end{eqnarray}

\begin{figure}
\begin{center}
\includegraphics[keepaspectratio,width=0.9\columnwidth]{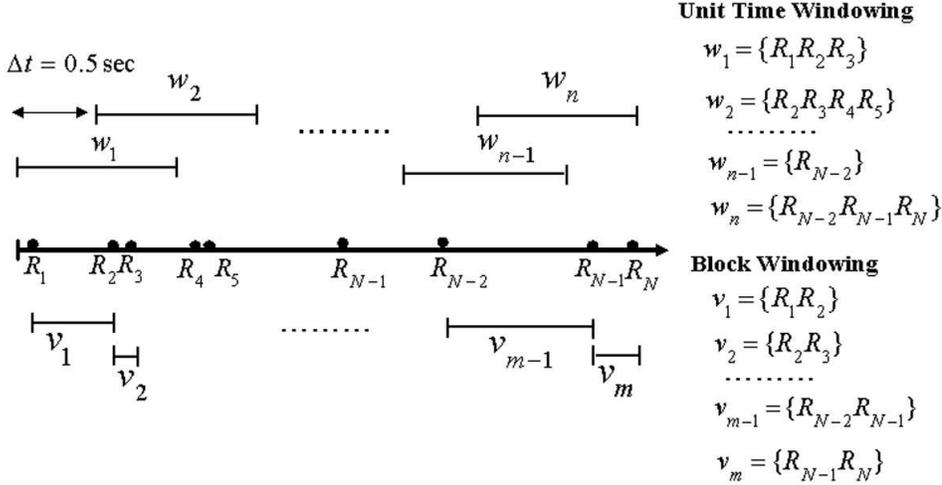}
\end{center}
\caption{\label{fig:fig2} Unit time windowing and block windowing method: Two different ways to compose a word sequence from a RR interval sequence. The upper case presents how to compose a word sequence from a RR interval sequence with the unit time windowing method, on the other hand, the lower case presents how to compose a word sequence with block windowing method. The former focus on unit time to define a word but the latter focus on the number of symbols, in this example, the block size is n=2, to define a word.} 
\end{figure}

A word $w_{i}$ composed by a unit time windowing method consists of n(i) symbols. The number of symbols, n(i), is different at each unit time scale $U_{T}$, satisfying $ \sum_{k=i}^{i+n(i)} x_{k} \leq U_{T} $. A symbol $s_{i+j}$, which is used to construct a word, is 1 for $|t_{i+j}|$ larger than $\tau$ and 0, otherwise. Here, $U_{T}$ is a unit time for constructing a word and $\tau$ is a RR interval threshold or a RR acceleration threshold for binary symbolization. Thus a word $w_{i}$ contains both information about the time scale and the event scale of the heart rate. Therefore, it defines a state of specific scale events of the cardiac system during a unit time. The event scale varies with 20 steps from 5\% to 95\% of the cumulative rank for all RR acceleration values from normal and pathologic HR data set. The time scale varies with 20 steps from 1 second to 10 seconds, which 1 second is sufficiently short to avoid an empty word.\\
Fig.\ref{fig:fig2} briefly illustrates how to compose a word sequence from a RR interval sequence. In the block windowing case, a word is defined regularly with the block size n=2 and the block is shifted to the one step next. In the unit time windowing case, a word is defined with a unit time $U_{T}$ and the unit time windows shifted by 0.5 second to define the next word, allowing the window overlap. The window shifting time, 0.5 second, is appropriately chosen to accumulate an enough number of sampled words. In this way, the unit time widowing method can match the number of words and the total measurement time of different RR intervals sets, while the block windowing method cannot. Through this procedure we can investigate the complexity of the symbolic RR sequence composed at each time and event scale combination.\\
With the above word constructing method, we can compose a word sequence from a RR interval sequence.
Let $ W=\{w_{1},w_{2}, \dots ,w_{n} \}$ be a finite, nonempty set which we refer to as an alphabet. A string or word over W is simply a finite sequence of elements of W. An alphabet W has the following relation,
\begin{eqnarray}
W^{n} \subseteq W_{n} \subseteq W^{\ast}, ~~ W^{0}=W_{0}=\Theta~~ \textrm{"empty word",}
\end{eqnarray}

where $W^{\ast}$ is the set of all words over W, including the empty word, $\Theta$, and $W_{n}$ the set of all words over W of length n or less and $W^{n}$ the set of all words over W of length n. The number of possible words of $W_{n}$ and $W^{n}$ are $N(n)=s^{1}+\dots s^{i}+ \dots s^{n}={s^{(n+1)}-1}/{s-1}$ and $N(n)=s^{n}$, respectively. And s is a s-ary number, which in this paper is binary(s=2). And the number of possible symbols in a word, $n(i)=Int(\frac{U_{T}}{ T_{s} } )$, varies with the unit time $U_{T}$, which is used to construct a word. $T_{s}$ is the smallest RR interval time and $n(U_{T})$ is an integer number. Since $n(U_{T})$ depends on $T_{s}$, which is sensitive on the ectopic beat or short term noisy beats, noise reduction should be done carefully in preprocessing. The $w_{i}$ word sequence composed by the unit time windowing method is a natural generalization of one based on the block windowing method.\\
%%%%%%%%%%%%%%%%%%%%%%%%%%%%%%%%%%%%%%%%%%%%%
%FIGURE2
 \\
%%%%%%%%%%%%%%%%%%%%%%%%%%%%%%%%%%%%%%%%%%%%%
Based on this unit time windowing method, we calculate the symbolic entropy, called the UTBE, which quantifies complexity of a word sequence at specific event and time scale region. 

The entropy, $H(U_{T},\tau)$, is a function of the time scale $U_{T}$ and the event scale $\tau$.
\begin{eqnarray}
H(U_{T},\tau)=-\sum_{i=1}^{N(U_{T})} p_{i} \log_{2} p_{i},
\end{eqnarray}
where $p(i)=\frac{N(w_{i})}{N(W)}$ is the estimation of the frequency of word $w_{i}$ and $N(w_{i})$ is the occurrence number of a word $w_{i}$ and $N(W)$ is the total number of sampled words. To calculate the exact probability of a word $w_{i}$, infinite words are considered with $N(W)\rightarrow \infty$. Then for s=2, UTBE varies between a lower bound and a upper bound at each time scale. The lower bound occurs for the completely regular case and the upper bound occurs for the completely random case where all words are equally probable.\\ 
In this paper, to construct an appropriate word ensemble from a RR sequence, the unit time scale is chosen up to 5 steps(about 2.8 seconds) for the UTBE. At this time scale, the largest number of symbols contained in a word from our data set is 8, which contribute to 2.9\% of the population. The number of possible words is $2^{8}=256$ and the number of sampled words from our RR interval sequence is 3,313 at this unit time scale. In the followings, we show that among two scales, the event scale is more significant for classification of healthy and two pathologic FHR groups than the time scale.\\

%%%%%%%%%%%%%%%%%%%%%%%%%%%%%%%%%%%%%%%%%%%%%%%%%%%%%%%%%%%%%%
% FIGURE4
\begin{figure*}
\begin{center}
\includegraphics[width=0.3\textwidth]{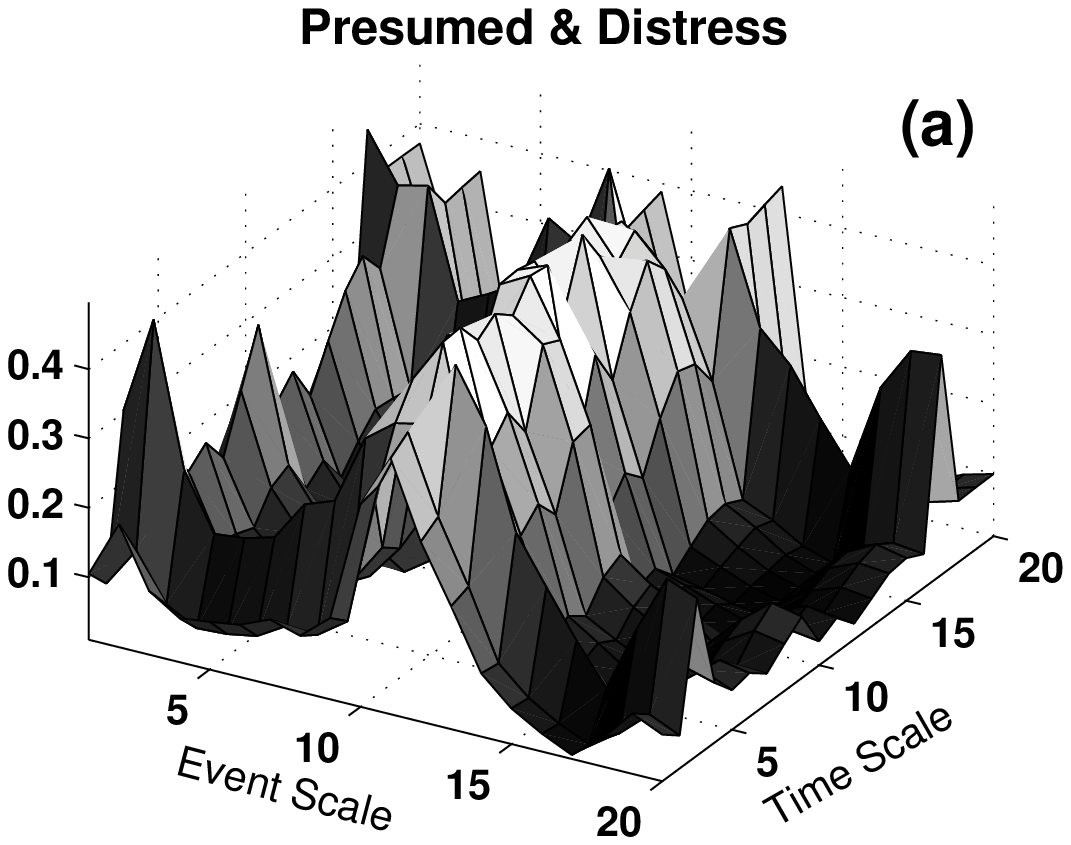}
\includegraphics[width=0.3\textwidth]{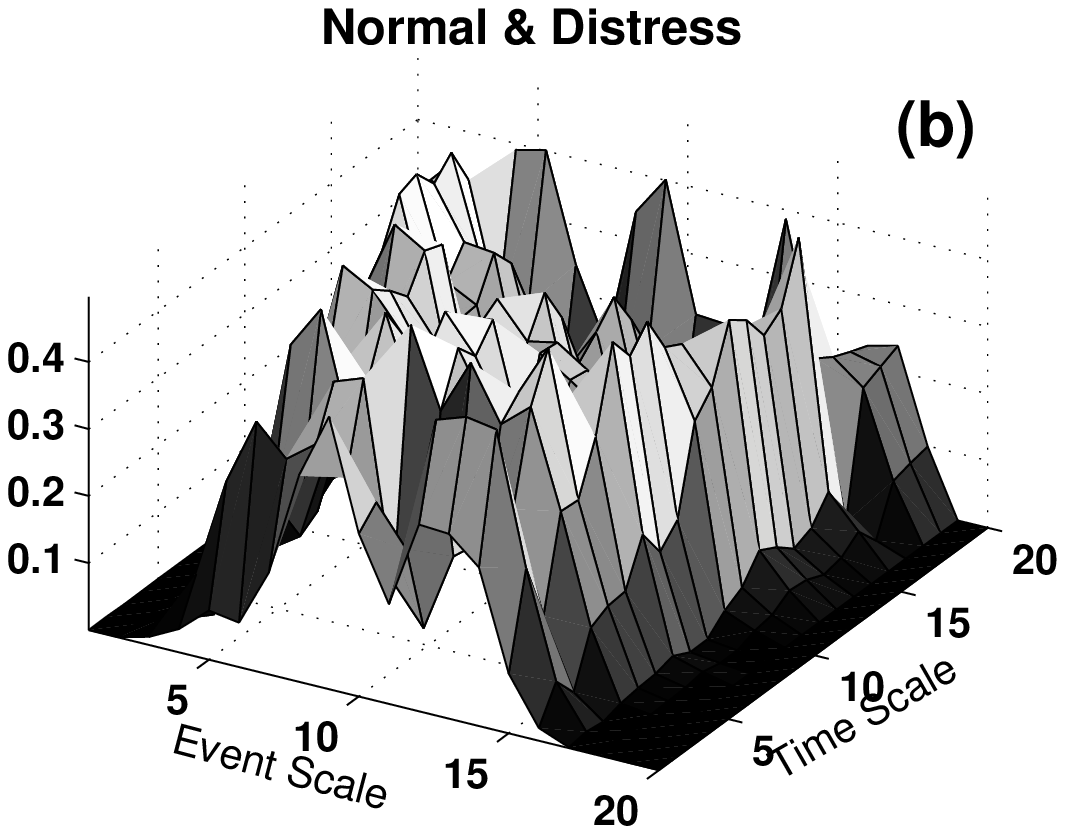}
\includegraphics[width=0.3\textwidth]{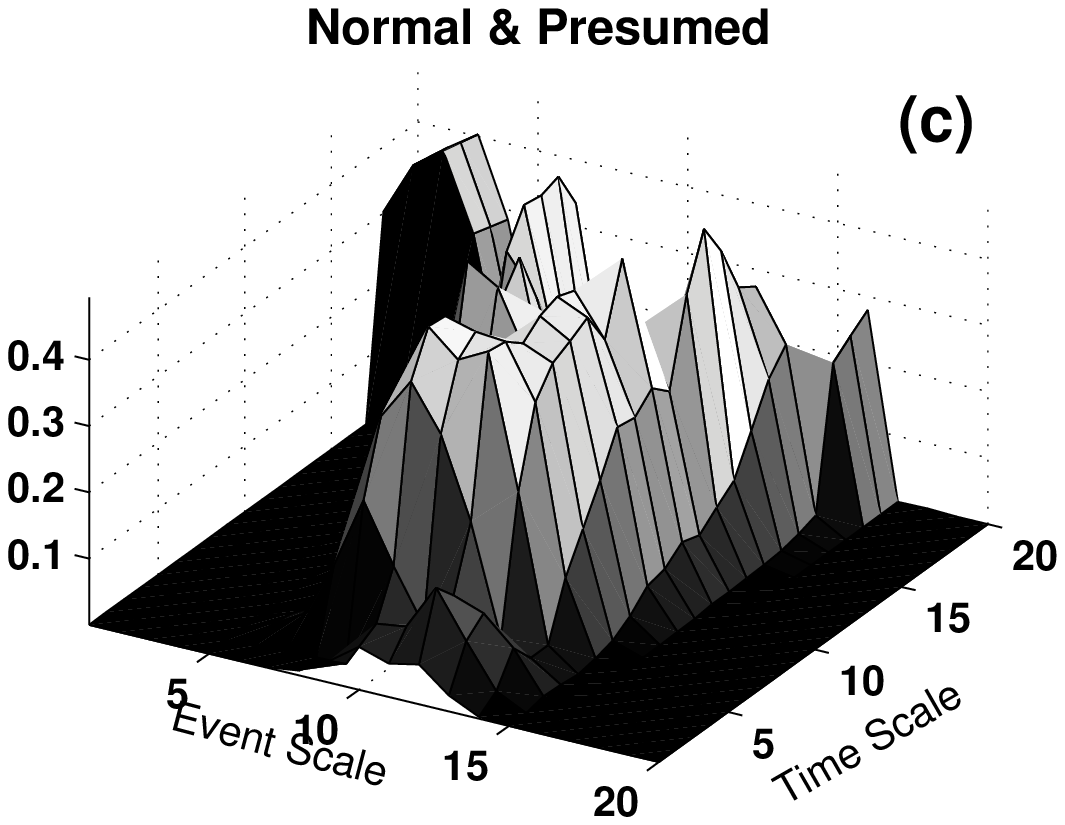}\\
\includegraphics[width=0.3\textwidth]{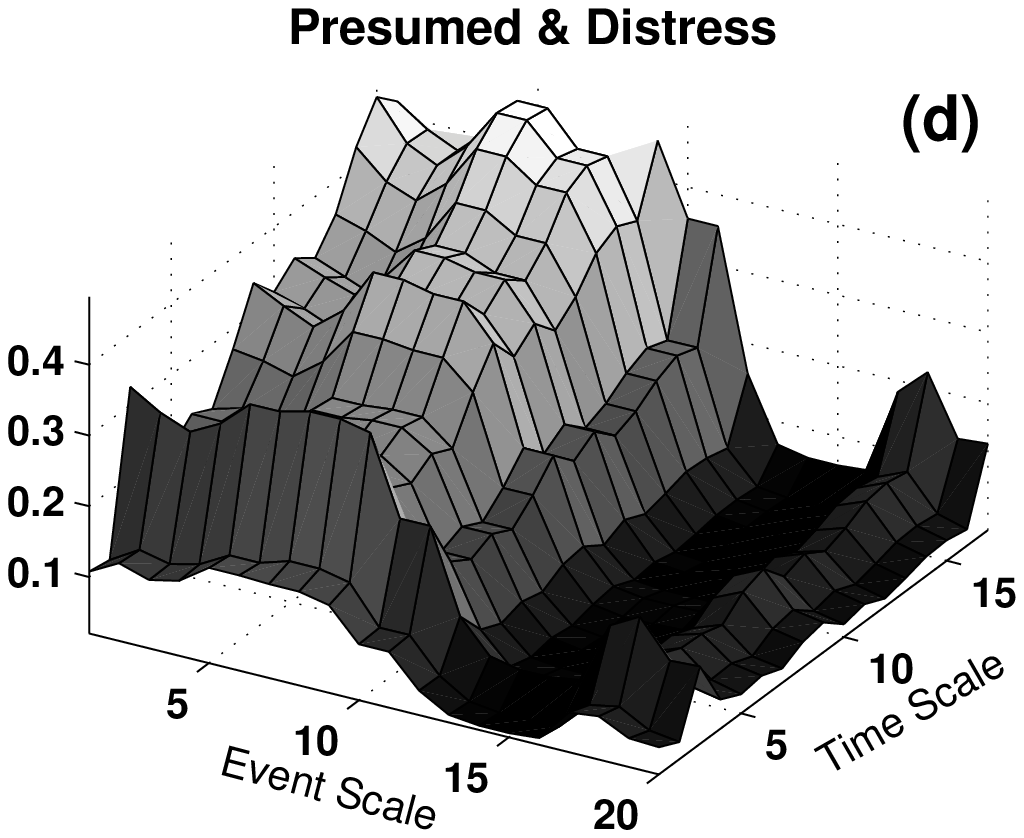}
\includegraphics[width=0.3\textwidth]{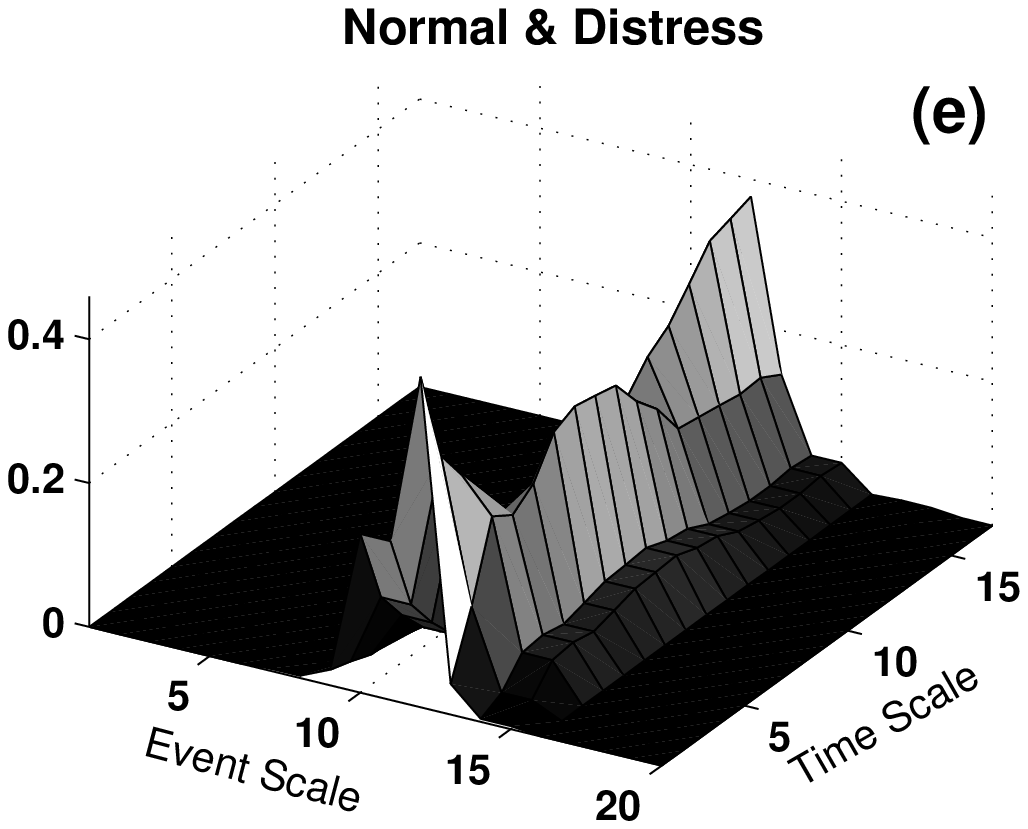}
\includegraphics[width=0.3\textwidth]{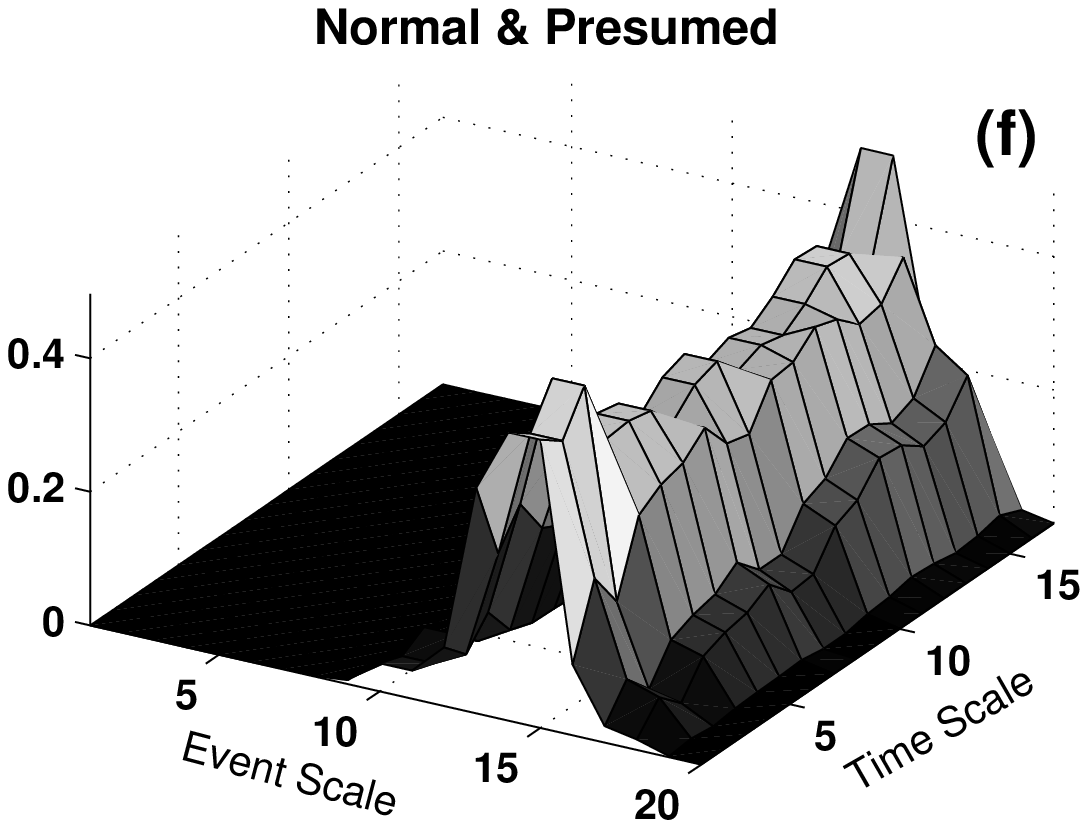}\\
\includegraphics[width=0.3\textwidth]{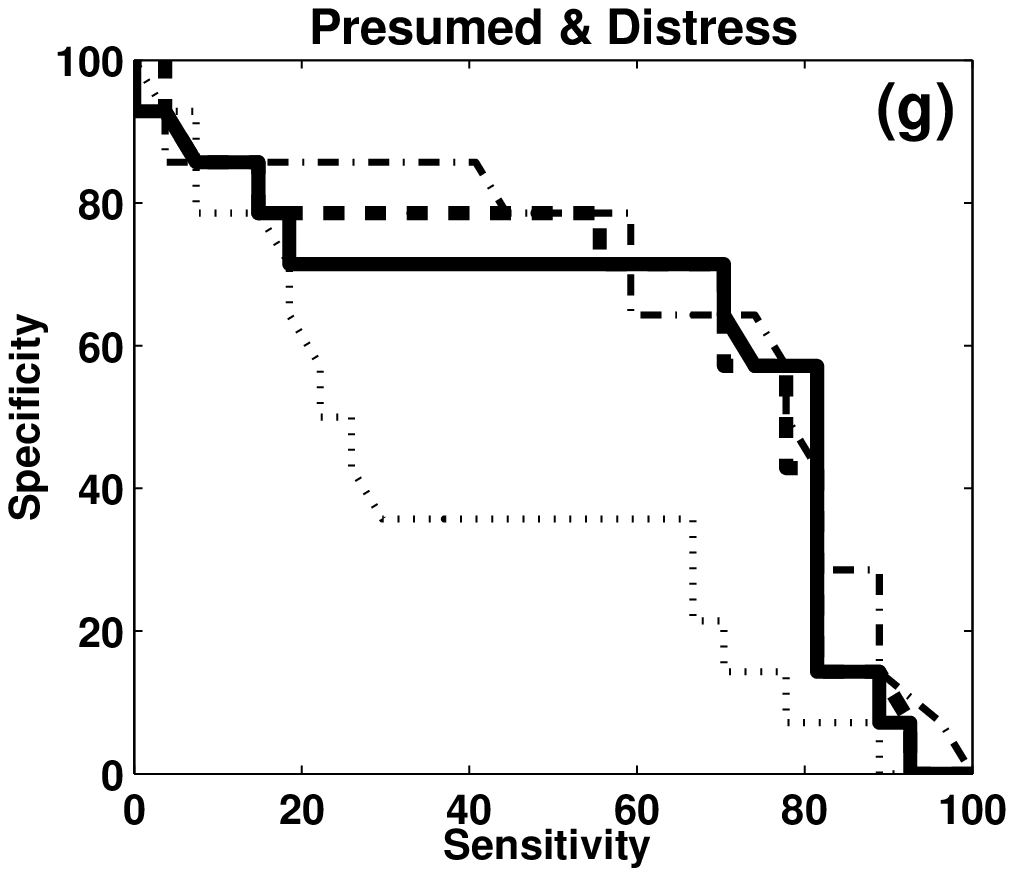}
\includegraphics[width=0.3\textwidth]{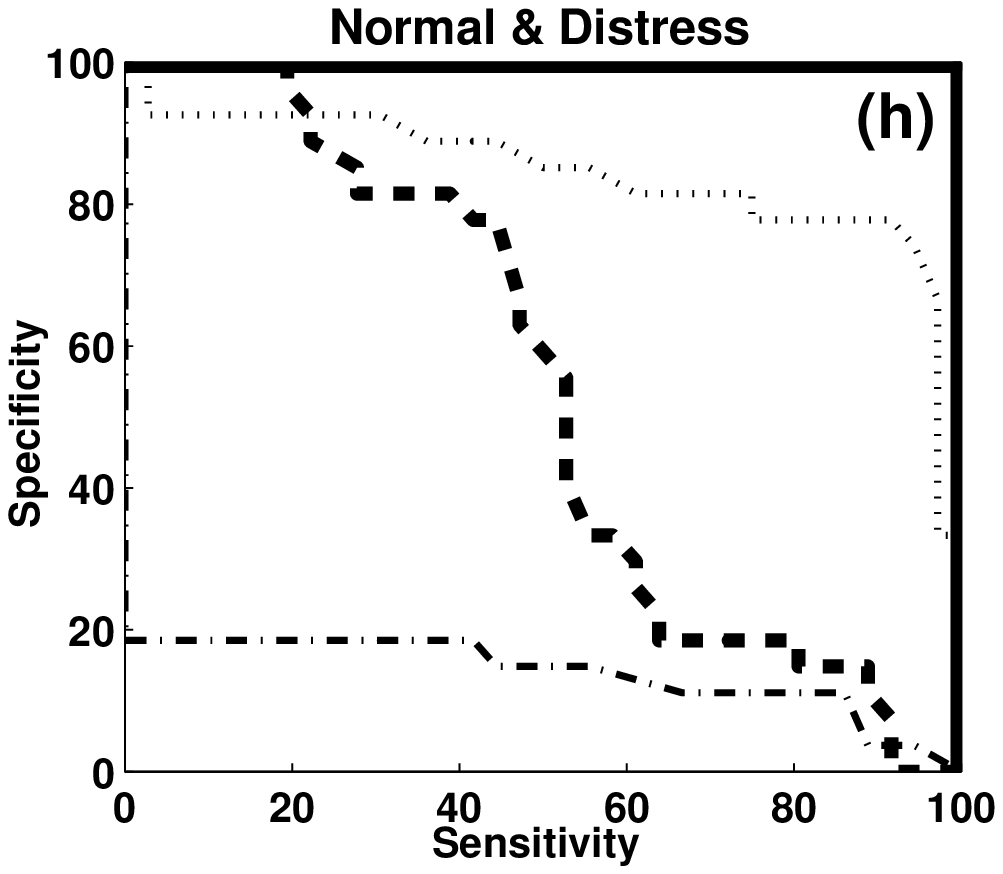}
\includegraphics[width=0.3\textwidth]{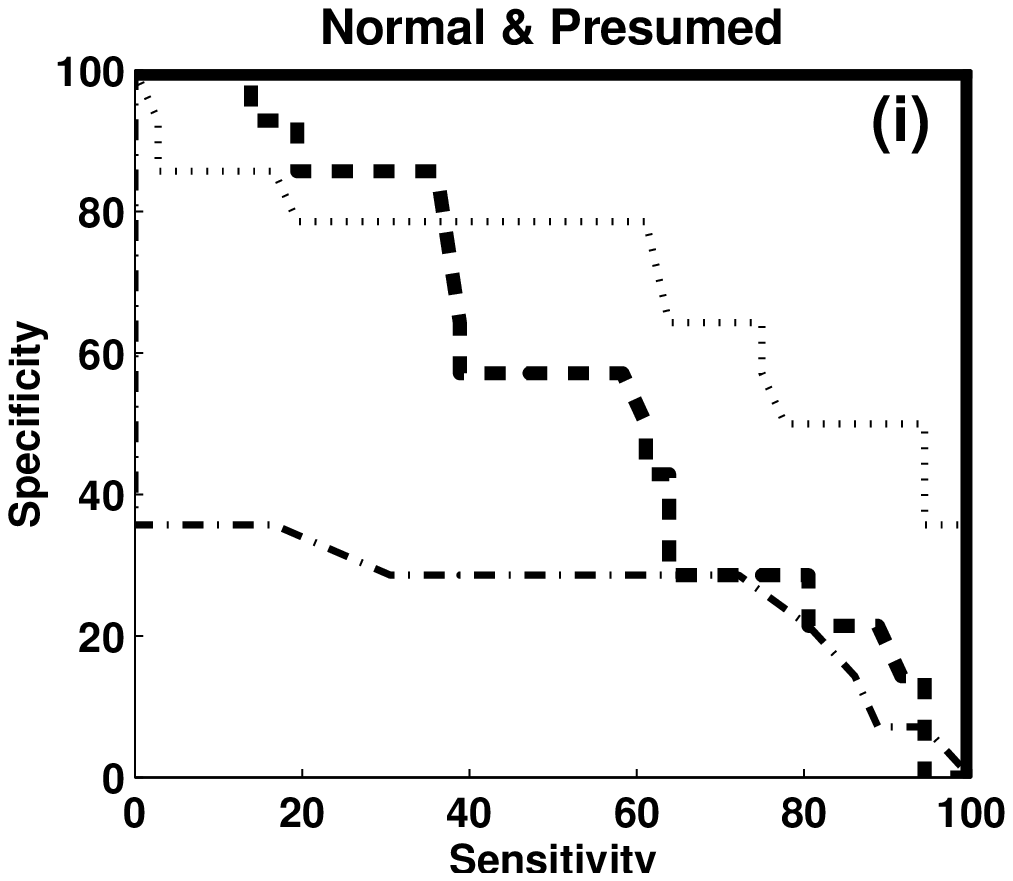}
\end{center}
\caption{\label{fig:fig4} Student t-test (p-value) and sensitivity \& specificity of the UTBE between the normal, the presumed distress and the acidotic distress groups using the RR interval and the RR interval acceleration as a symbolization threshold. (a-c) The p-values between three groups in the ($U_{T}, \tau$) parameter space using the RR interval as a threshold. (d-f) The p-values between three groups in the ($U_{T},\tau$) parameter space using the RR interval acceleration as a threshold.[The presumed and acidotic distress, the normal and the presumed distress, the normal and the acidotic distress]. The RR interval acceleration as a symbolization threshold gives better performance than that of RR interval in distinguishing between the normal, the presumed distress and the acidotic distress groups.
(g-i) The sensitivity \& specificity between three groups in two UTBEs and two linear properties(mean, standard deviation).( dotted line: the mean , dot dashed line : the standard deviation, bold-solid line : UTBE using the RR interval acceleration, bold-dotted line : UTBE using the RR interval), (g) the presumed and the acidotic distress, (h) the normal and the presumed distress, (i) the normal and the acidotic distress. In these figures, the best performing sensitivities and specificities are presented for each case.}
\end{figure*}
%%%%%%%%%%%%%%%%%%%%%%%%%%%%%%%%%%%%%%%%%%%%%%%%%%%%%%%%%%%%%%
\subsection{\label{sec:level2} Application of the UTBE method} 
In Fig.4, we applied the UTBE method to the three FHR groups and show the result of the search for all event and time scale regions. In order to compare UTBE distributions between the normal and two pathologic groups, all p-values of the student t-test in the $(U_{T}, \tau)$ parameter plane are presented in Fig. 4(a-f). Since the scale characteristics of three groups are not known a priori, we scanned all event and time scales for optimization of the classification. For the presumed distress and the acidotic distress groups in the Fig.\ref{fig:fig4}(a) and Fig.\ref{fig:fig4}(d), the event sequences composed by only the 50-80\% cumulative rank region in the RR interval acceleration are significantly distinguished over all time scales($p<0.01$). This suggests that the presumed distress and the acidotic distress groups have relatively different characteristics scale complexity on this specific RR interval acceleration region. These two groups could not be discerned by their linear properties[the p-value of (mean/standard deviation):(0.28/0.903)]. For the normal and the acidotic distress groups in Fig.\ref{fig:fig4}(b) and Fig.\ref{fig:fig4}(e), the event sequences composed by most RR interval acceleration scales except the 50-75\% cumulative rank region are distinguished in their complexity($p< 10^{-4}$), while in event sequences composed by only narrow RR interval scale regions, with 5\% and 95\% cumulative rank ones, show the difference. It suggests that the RR interval acceleration of the normal and the acidotic distress group contains significant information on different internal dynamics of the cardiac system, while the RR interval does not. For the normal and the presumed distress groups in Fig.\ref{fig:fig4}(c) and Fig.\ref{fig:fig4}(f), most RR interval scales could not discriminate these two groups except narrow regions, with 5-35\% ranks and 90-95\% cumulative ranks, while a wide RR acceleration scale region below the 50\% cumulative rank significantly discriminates these two groups($p<1.1\times 10^{-4}$).\\
The above results suggest that the normal, the presumed distress and the acidotic distress groups exhibit characteristic scales with typical dynamics in those scale regions. The information on the characteristic event and time scale regions of three different groups can be used to determine the optimal parameters for the classification of different group. Between two types of scales, the event scale is more effective than the time scale in classifying these groups. When two groups are discriminated at an event scale region, these groups are also well discriminated over most time scales. These results suggest that the RR interval acceleration contains typical information about cardiac dynamics of three groups, whereas the RR interval does not. Since slow or fast acceleration of the cardiac system is associated with fetal vagal activity and the motherly-fetal respiratory exchange system, it may provide some clues to which functional difference of cardiac systems causes such difference between healthy and pathological groups.\\
In Fig.\ref{fig:fig4}(g)-(i), the best sensitivity and specificity in the $(U_{T},\tau)$ parameter plane is presented for the four measures of statistics and complexity for comparison, respectively.\\
The sensitivity determines that when a UTBE value is given for classifying two groups(A,B) as a threshold, what percentage of the subjects involved in the group A is correctly classified by the given threshold UTBE value. The specificity determines that when a UTBE value is given for classifying two groups as a threshold, what percentage of the subjects involved in the other group B is correctly excluded from the group A. So if the sensitivity and specificity are all $100\%$, the two groups are completely classified by the given threshold. Here, the best sensitivity and specificity are achieved after calculating them at all points of the ($U_{T},\tau$) parameter space, varying the threshold from the minimum UTBE to the maximum UTBE value of the calculated UTBE set\cite{Greenhalgh1997}. Here, the best sensitivity and specificity is determined as the highest values along the diagonal in the plane of the sensitivity and specificity.
As a result, UTBE using the RR interval acceleration as a threshold provides the best performance in classification of the presumed distress and the acidotic distress, the normal and the acidotic distress, and the normal and the presumed distress groups. Surprisingly, for the normal group and two types of distress groups, UTBE using the RR interval acceleration as a threshold completely discriminates these groups with sensitivity 100\%, and the specificity 100\%, which could not be archived with two linear properties and the UTBE using the RR interval threshold. For the presumed distress and the acidotic distress groups, both UTBEs lead to the same result(sensitivity=71.4\%, specificity=72\%).
%%%%%%%%%%%%%%%%%%%%%%%%%%%%%%%%%%%%%%%%%%%%%%%%%%%%%%%%%%%%%%%%%%%%%%%%%%%%%%
% FIGURE5
\begin{figure}
\begin{center}
\includegraphics[width=0.45\textwidth]{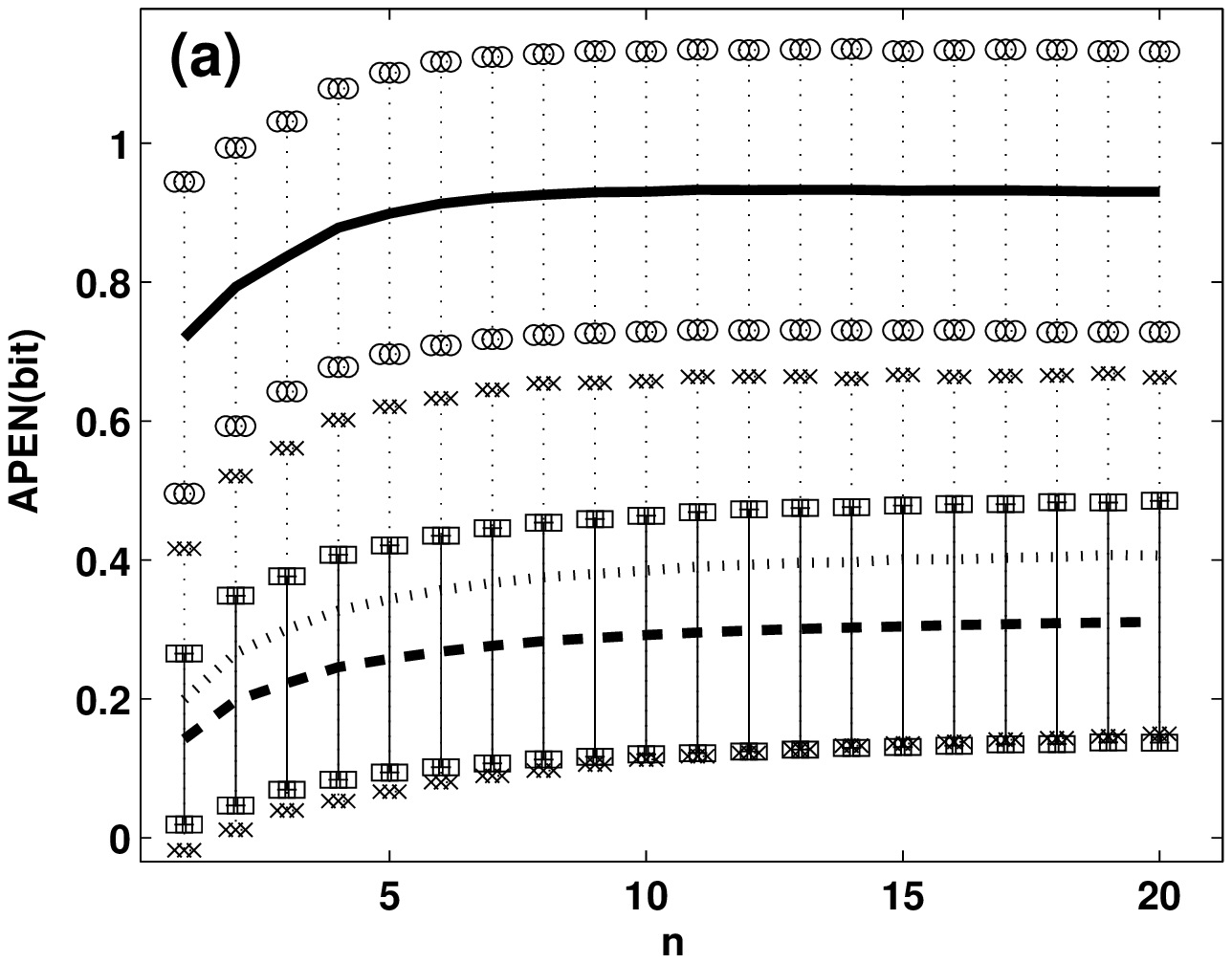}
\includegraphics[width=0.45\textwidth]{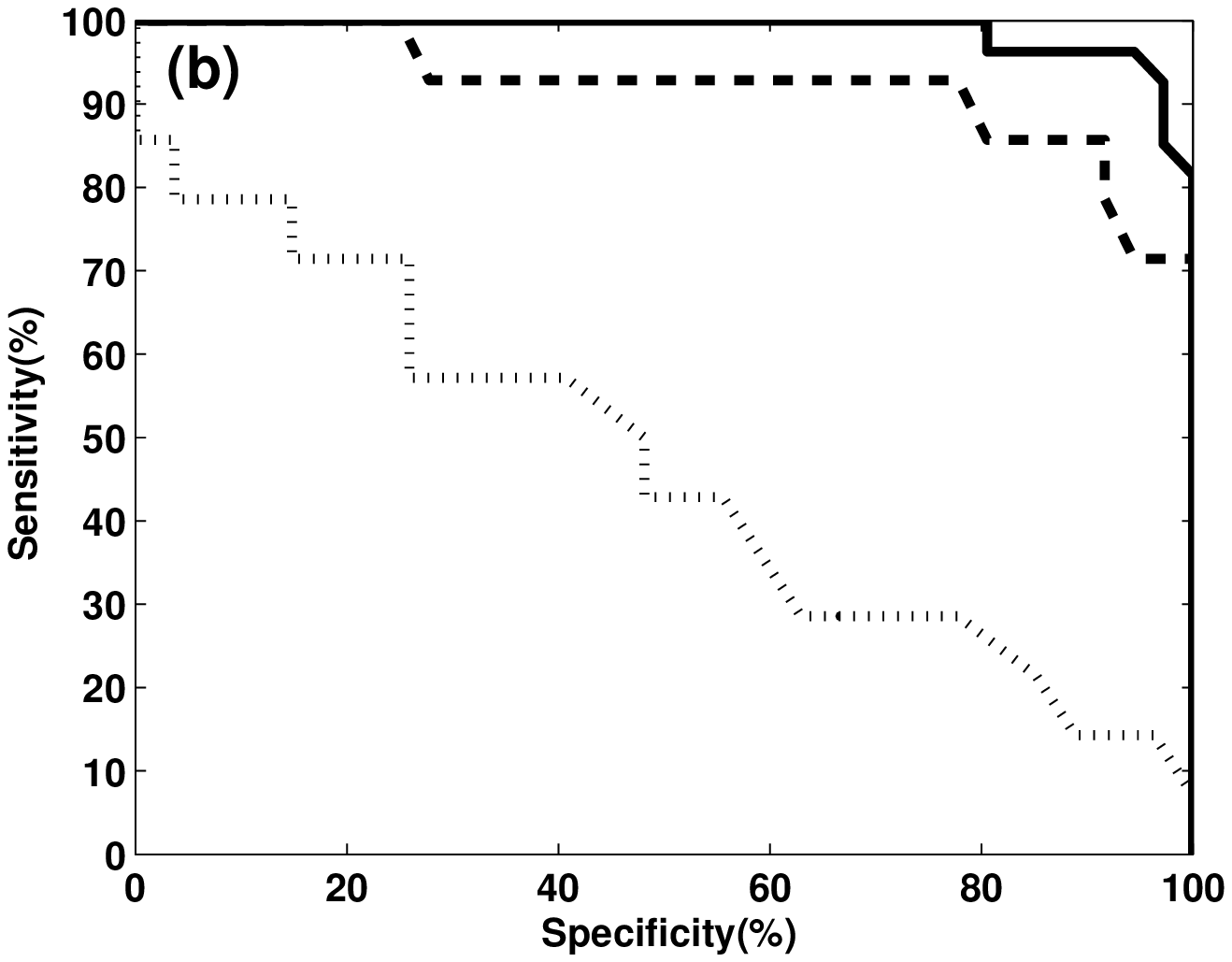}
\end{center}
\caption{\label{fig:fig5} (a) The APEN with error bar for different FHR groups. The normal(bold solid line) has the highest complexity and the presumed(dotted line) and the acidotic FHR(bold dotted line) have the lower APEN values at each time scale. (b) The best sensitivity and specificity of all FHR groups. The best sensitivity and specificity is (94\%, 95\%) for the normal group and the presumed distress group(solid line), (85\%,86\%) for the normal group and the acidotic distress group(bold dotted line) and (48\%,48\%) for the presumed group and acidotic distress group(dotted line).}
\end{figure}
%%%%%%%%%%%%%%%%%%%%%%%%%%%%%%%%%%%%%%%%%%%%%%%%%%%%%%%%%%%%
\subsection{\label{sec:level4} Comparison with the conventional method(MultiScale Entropy)}
In this section, we compare the performances of the UTBE and the MultiScale Entropy(MSE), which has been widely used in the hear rate analysis\cite{Costa2002,Chialvo2002,Costa2002_2,Costa2003,Costa2005}. The Multi-Scale Entropy calculates the approximate entropy(APEN) or the sample entropy at different scales with heart rate data, which measures the regularity of a given data. Since the same data length(N=4061) is used in order to remove the dependency on the data length, the approximate entropy is chosen instead of the sample entropy. First, in order to calculate the APEN, the RR sequence of length N(=4061) is divided into segments of length n and the mean value is calculated for each segment. With the coarse-grained sequence at each scale n, the APEN is computed with the following, the parameters; the embedding dimension m=2 and the delay $\tau=1$\cite{Costa2002,Costa2005}. In the calculation of the Multi-Scale Entropy, we use two types of r values; one is determined from the original data at n=1[$r=0.15\times SD(n=1)$] and the other is variable to be determined at all n scales[$r(n)=0.15\times SD(n)$]. By using the variable r(n) we can remove the effect of variation due to the coarse-graining process, where SD(n) denotes the standard deviation of the coarse-grained sequence at a scale n\cite{Nikulin2004}. Since the Multi-Scale Entropy for two cases lead to the similar results, we present here one for the first case. In Fig.5(a) and (b), we present the performance of Multi-Scale Entropy in the classification of three fetal heart rate groups. In Fig 5.(a), the best p-values for each pair of groups in the student t-test are $p=1.8\times 10^{-10}(n=4)$ for the normal vs the acidotic distress, $p=10^{-10}(n=4)$ for the normal vs the presumed distress and p=0.143(n=1) for the presumed distress vs the acidotic distress. This result shows that the mean Multi-Scale Entropy values of each group are significantly different between the normal and two pathologic groups, but the mean Multi-Scale Entropy values of two pathologic groups are not distinguishable. In order to check the possibility of classification, we investigate the sensitivity and specificity as in the UTBE. Fig.5(b) presents the best sensitivities and specificities selected from the calculation in all scales. The best sensitivity and specificity is $(94\%,95\%)$ at the scale(n=4) of the normal and the presumed distress case, $(85\%, 86\%)$ for the normal and the acidotic distress case and $(48\%,48\%)$ for the presumed distress and the acidotic distress case. The classification performance of Multi-Scale Entropy is not better than that of the UTBE in all classification cases from the three groups.
This is because UTBE searches all the event and time scales to find the optimal classification of different characteristics of healthy and two pathologic data, while Multi-Scale Entropy searches only the time scale.

\section{\label{sec:leve14} Scale Characteristics of Healthy and Pathologic FHRs.}
%%%%%%%%%%%%%%%%%%%%%%%%%%%%%%%%%%%%%%%%%%%%%%%%%%%%%%%%%%%%%%%%%%%%%%%%%%%%%
% FIGURE6
\begin{figure}
\begin{center}
\includegraphics[width=0.45\textwidth]{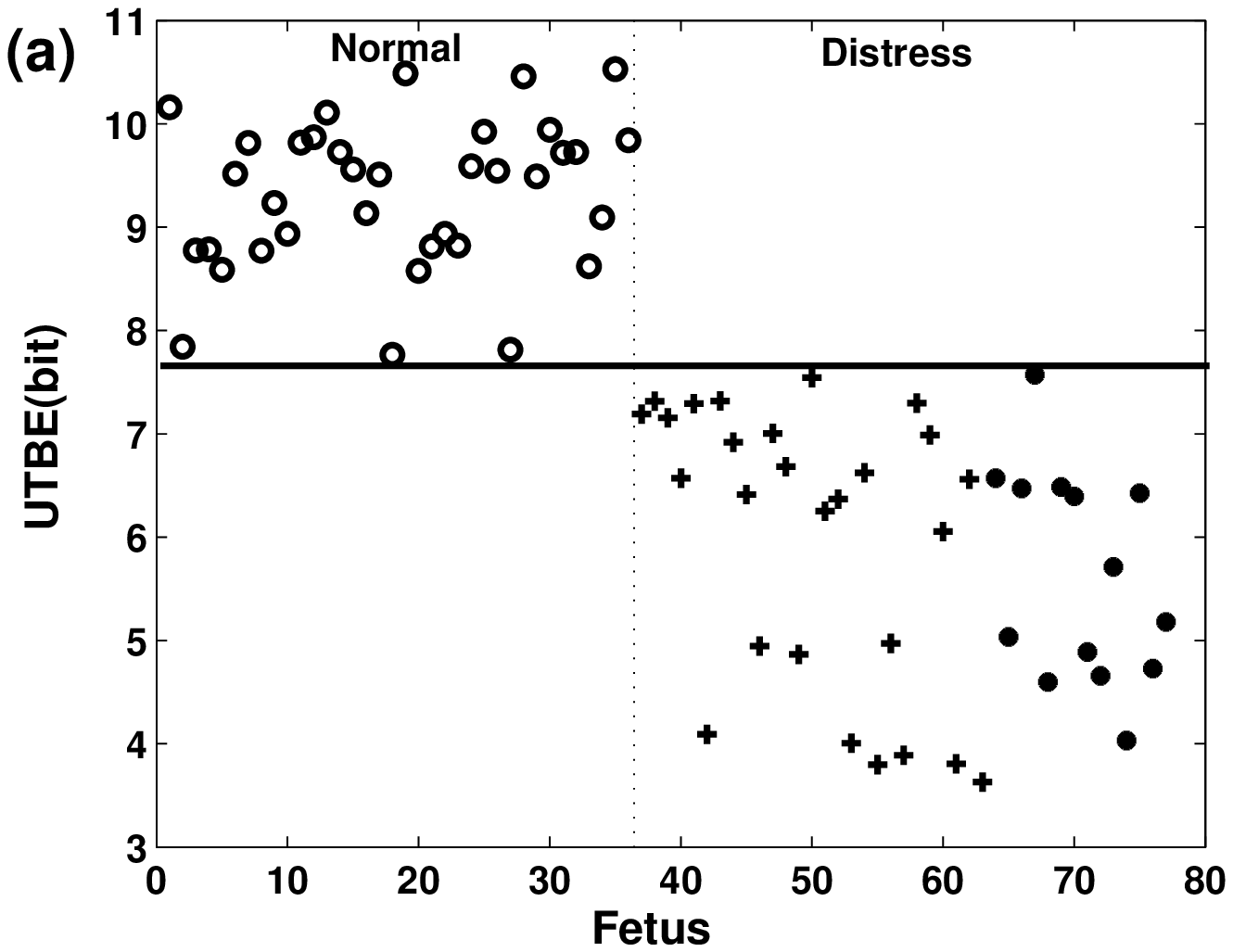}
\includegraphics[width=0.45\textwidth]{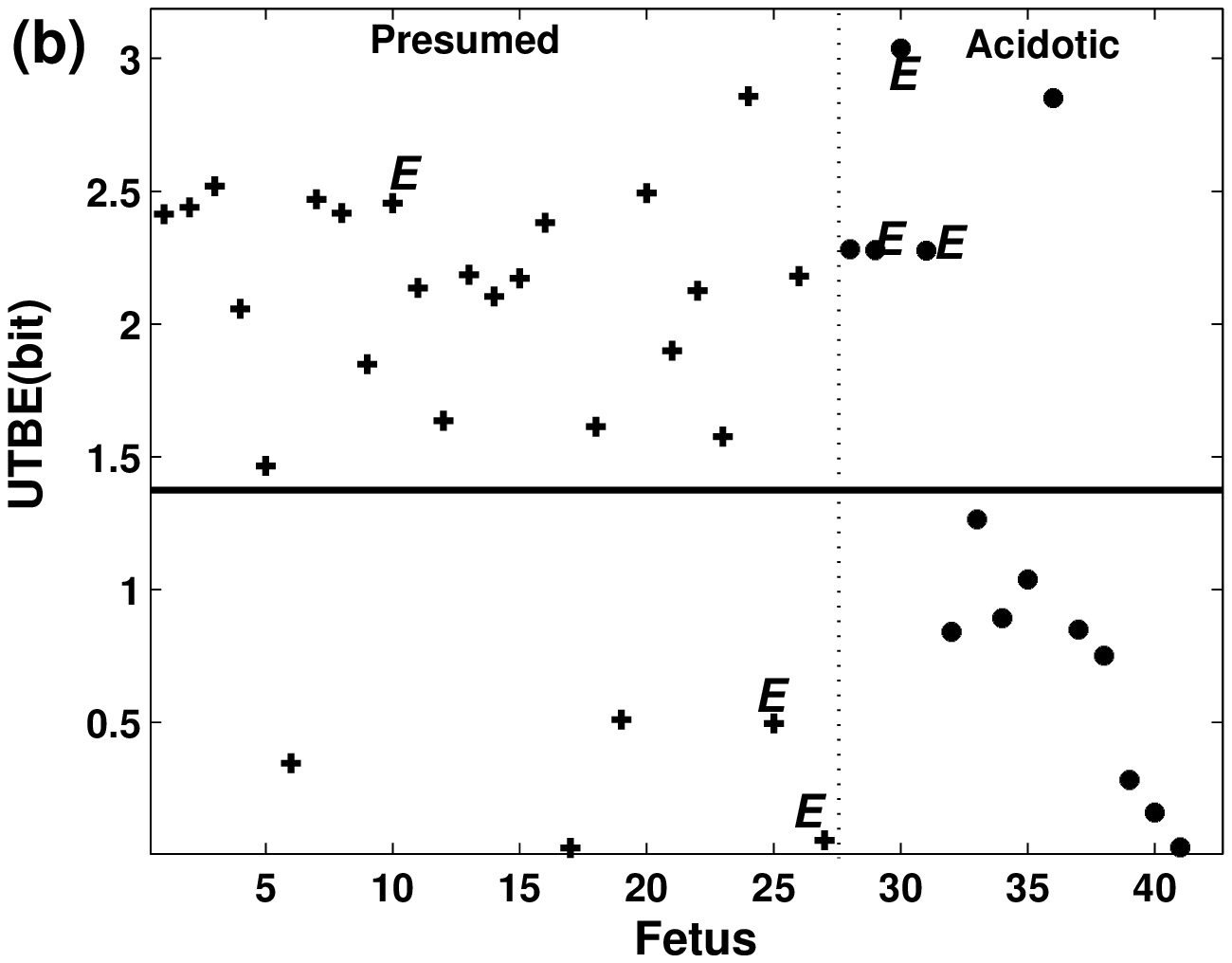}
\end{center}
\caption{\label{fig:fig6} (a) The normal and the pathologic FHR groups are distinguished at a specific time and event scale(3sec,5\% rank). Here, the two groups are divided by the threshold(UTBE=7.8bit). (b) The presumed FHR and the acidotic FHR groups are divided by the threshold(UTBE=1.4). The marker "E" indicates the three fetuses in each group, whose umbilical artery PHs are closest to 7.15.}
\end{figure}
%%%%%%%%%%%%%%%%%%%%%%%%%%%%%%%%%%%%%%%%%%%%%%%%%%%%%%%%%%%%
In this section, using the difference in the characteristic event scale between the normal and two pathologic groups, we distinguish the normal, the presumed and acidotic groups systematically. In Fig.4(e) and (f), the normal group is significantly distinguished from the presumed distress group and the acidotic distress group in the lower event scale region. In Fig.4(d), the presumed distress group and the acidotic distress group are distinguished in the relatively higher event scale region. Thus, with these characteristics, we can make a strategy to systematically differentiate these groups based on the UTBE. First, as in Fig.6(a), we separate the normal and pathologic groups by a threshold of 7.8 in the UTBE, which is determined as the smallest UTBE value of the normal group, at the specific time and event scale(3 seconds,$ 5\%$ rank),respectively. From this procedure, we can differentiate the normal and the pathologic groups with a 100\% accuracy. Then, for the fetuses deviating from the normal group, usually having the lower UTBE, we try to separate the presumed and the acidotic distress groups. In this case, we determine the threshold UTBE as 1.4 bits in the time and event scale, which are the first step(1 sec) and the tenth step(about 50\% rank), respectively. These time and event scales are selected from the scan of the parameter space as in Fig.4(d), in which these two groups are distinguished well on the scale region above the tenth event scale. With this threshold, 9 acidotic fetuses out of 14 are distinguished. However, some ambiguity still remains. The clinical determination of the presumed distress fetus and the acidotic distress fetus was carried out by the umbilical artery PHs. In Fig.6(b), we mark the three fetuses with 'E' in each group, who have the umbilical artery PHs closest to the threshold value of PHs, 7.15. The range of umbilical artery PHs measured is from 6.863 to 7.38. The PHs values of the six fetuses are 7.229, 7.24 and 7.226 in the presumed distress group and 7.15, 7.16 and 7.144 in the acidotic distress group. Since all the pathologic fetuses were delivered by caesarean surgery after several signs of distress(severe variable, later deceleration, bradycardia, or tachycardia), it is not certain if all the acidotic distress fetuses would go to the emergency state or all the presumed distress would be in the safety state. Therefore, if the fetuses marked with 'E' are excluded in this analysis, the characteristic of two groups can be clarified more clearly. As a result, most acidotic distress fetuses are less complex than the most presumed distress fetuses at the time and event scale regions. In the comparison of three groups, two pathologic groups are less complex than the normal group, while in the comparison of two pathologic groups the acidotic distress group is less complex than the presumed group. We find that in order to distinguish the normal and pathologic fetuses a small event scale(5\% rank) and a relatively large time scale(3 sec) of fetal heart dynamics is useful. On the other hand, in order to distinguish the presumed distress and the acidotic distress fetuses a large event scale(about 50 \% rank) and a relatively small time scale(1 sec) of fetal heart dynamics is more appropriate. With these scale regions, we were able to reduce the wrong diagnosis rate from $63.4\%(26/41)$ to $24.3\%(10/41)$.\\

%%%%%%%%%%%%%%%%%%%%%%%%%%%%%%%%%%%%%%%%%%%%%%%%%%%%%%%%%%%%%%%%%%%%%%%%%%%%%%%
\section{CONCLUSION}
In this paper, we investigated the event and time scale structure of the normal and pathologic groups. We also introduced and calculated the UTBE method in the appropriate event and time scale region to distinguish the three groups. To extract meaningful information from the data set, the scale structure of the RR interval acceleration is found to be more helpful than that of the RR interval. In the comparison of the UTBE over all event and time scale regions, we found that the normal, the presumed distress and the acidotic distress groups have relatively different event scale structures in the RR interval acceleration. In particular, for the normal and two pathologic groups the UTBE from the RR acceleration threshold completely classifies these groups in a chosen scale regions, although both linear properties and UTBE using the RR interval threshold performs worse. In the case of the presumed distress and the acidotic distress groups, it also provides better classification performance than other measures. The comparison with the Multi-Scale Entropy also shows that the UTBE method performs better. It is due to the fact that the UTBE approach searches the event and time scale region, while the Multi-Scale Entropy method searches only the time scale.\\ 
Based on the difference in the scale structure, we are able to systematically distinguish three FHR groups. The normal and the pathologic groups are separated in a small event scale(5\% rank) and a large time scale(3 sec). Then, the fetuses deviating from the normal group are separated into the presumed fetuses and the acidotic fetuses at a relatively large event scale(50\% rank) and a small time scale(1sec). From these scale regions, we reduce the wrong diagnosis rate significantly. \\
The results suggest that the UTBE approach is useful for finding the characteristic scale difference between healthy and pathologic groups. In addition, we can make a more reliable comparison between all fetuses by simultaneous matching of the measurement time and the number of words. This approach can be applied to the other unevenly sampled data taken from complex systems such as biomedical, meteorological or financial tick data. In this study, we also reconfirm that the more pathological a fetus is the less complex its dynamics, following the pathological order, from the acidotic distress, the presumed distress and to the normal fetus. This indicates that in spite of the peculiarity of the fetal cardiac system, the generic notion of the complexity loss can be applied to the fetal cardiac system.\\
But these results come from a retrospective test under the well elaborated condition. In a practical view point, the prospective test is necessary to confirm the selected scale regions and its classification performance. We will further test these results with more subjects for the purpose of practical application of this analysis method.

$~~~$ We thank Dr. Sim and Dr.Chung in Dep. Obstetrics and Gynecology, Medical College, Catholic Univ. for their assistance on clinical data and helpful comments. This work has been supported by the Ministry of Education and Human Resources through the Brain Korea 21 Project and the National Core Research Center Program. 


\begin{thebibliography}{99}
\bibitem{Altimiras1999} J. Altimiras, Comp. Biochem. Physiol. A \textbf{124},447 {(1999)}
\bibitem{Havlin1999} S. Havlin, S. V. Buldyrev, A. Bunde \textsl{et al.}, Physica A \textbf{273}, 46 {(1999)}
\bibitem{Costa2002} M. Costa, A. Goldberger and C.-K. Peng, Phys. Rev. Lett. \textbf{89}, 068102 {(2002)}
\bibitem{Chialvo2002} D. R. Chialvo, Nature \textbf{419}, 263 {(2002)}
\bibitem{Costa2002_2} M. Costa, A. Goldberger and C-K. Peng, Comput. Cardiol. \textbf{29}, 137 {(2002)}
\bibitem{Costa2003} M. Costa and J. A. Healey, Comput. Cardiol. \textbf{30}, 705 {(2003)}
\bibitem{Costa2005} M. Costa and A. Goldberger and C.-K. Peng, Phys. Rev. E \textbf{71}, 021906 {(2005)}
\bibitem{Wood1999} C. E. Wood and H. Tong, Am. J. Physiol. \textbf{277}, R1541 {(1999)}
\bibitem{Lee2005} U. C. Lee, S. H. Kim and S. H. YI, Phys. Rev. E\textbf{71}, 061917 {(2005)}
\bibitem{Magenes2002} G. Magenes, M. G. Signorini and D. Arduini, Proc. the Second Joint EMBS/BMES Conference, October 23-26, 2002
\bibitem{Magenes2003} G. Magenes, M. G. Signorini and M. Ferrario \textsl{et al.}, Proc. the 25th Annual International Conference of the IEEE/EMBS, Sep. 17-21, 2003
\bibitem{Signorini2003} M. G. Signorini, G. Magenes and S. Cerutti\textsl{et al.}, IEEE Trans. Biomed. Eng. \textbf{50}, 365 {(2003)}
\bibitem{Gomes2002}  M. E. D. Gomes, H. N. Guimaraes, A. L. P. Ribeiro and L. A. Aguirre, Comput. Biol. Med. \textbf{32}, 481 {(2002)}
\bibitem{Greenhalgh1997} The sensitivity is a true positive rate defined as $ \frac{A}{A + C}$, and the specificity is a true negative rate defined as $ \frac{D}{B + D}$. (A: true positive, B: false positive, C: false negative, D: true negative). More details are available at http://bmj.bmjjournals.com/ or see T. Greenhalgh, BMJ \textbf{315}, 540 {(1997)}   
\bibitem{Nikulin2004} V. V. Nikulin and T. Brismar, Phys. Rev. Lett. \textbf{92}, 089803 {(2004)}
\end{thebibliography}
\end{document}